# Thermodynamic modeling with uncertainty quantification using the modified quasichemical model in quadruplet approximation: Implementation into PyCalphad and ESPEI

Jorge Paz Soldan Palma,[a] Rushi Gong,[a] Brandon J. Bocklund,[a,b] Richard Otis,[c] Max Poschmann,[d] Markus Piro,[d] Tatiana G. Levitskaia,[e] Shenyang Hu,[e] Nathan Smith,[a] Yi Wang,[a] Hojong Kim,[a] Zi-Kui Liu,[a] and Shun-Li Shang,[a,*]

[a] Department of Materials Science and Engineering, The Pennsylvania State University, University Park, PA 16802, USA

[b] Lawrence Livermore National Laboratory, Livermore, CA 94550, USA

[c] Engineering and Science Directorate, Jet Propulsion Laboratory, California Institute of Technology, Pasadena, CA 91109, USA

[d] Faculty of Energy Systems and Nuclear Science, Ontario Tech University, Oshawa, ON, Canada

[e] Pacific Northwest National Laboratory, PO Box 999, Richland, WA 99352, USA

*Corresponding author. E-mail address: sus26@psu.edu

Abstract:

The modified quasichemical model in the quadruplet approximation (MQMQA) considers the first- and the second-nearest-neighbor coordination and interactions, particularly useful in describing short-range ordering in complex liquids such as molten salts, slag in metal processing, and electrolytic solutions. The present work implements the MQMQA into the Python based open-source software PyCalphad for thermodynamic calculations. This endeavor facilitates the development of MQMQA-based thermodynamic database with uncertainty quantification (UQ) using the open-source software ESPEI. A new database structure based on Extensible Markup Language (XML) is proposed for ESPEI evaluation of MQMQA model parameters. Using the KF-$NiF_2$ system as an example, we demonstrate the successful implementation of MQMQA in PyCalphad through thermodynamic calculations of Gibbs energy, equilibrium quadruplet fractions, and phase diagram, as well as database development with UQ using ESPEI. The present implementation offers an open-source capability for performing CALPHAD modeling for complex liquids with short-range ordering using MQMQA.

# 1 Introduction

Computational thermodynamics via the CALPHAD modeling approach [1–3] is a key methodology to understand, predict, and optimize materials in terms of thermodynamic properties of individual phases – the building blocks for the materials of interest [4,5]. For example, thermodynamics is needed to understand critical features of molten salts used in the Molten Salt Reactors (MSRs) [6], optimize fuel composition, and define safety margins in order to fulfill multiple requirements of molten salts such as i) low-melting temperature to decrease the risk of freezing and precipitation, ii) low vapor pressure at operating temperatures, iii) high heat capacity to store more energy, iv) thermodynamic stability, v) good solubility of fissile and fertile isotopes (e.g., $^{233}$U, $^{235}$U, $^{238}$U, and/or $^{239}$Pu), and vi) superior corrosion resistance [7,8].

Thermodynamic description of complex liquids of molten salts is based mainly on the two-sublattice ionic liquid model [9] or the modified quasichemical model (MQM) [10,11]. In the CALPHAD community, the ionic model is implemented in software tools such as Thermo-Calc [12], FactSage [13], Pandat [14], OpenCalphad [15], and PyCalphad/ESPEI [16,17]; while the MQM and especially the MQM in the quadruplet approximation (MQMQA) models are implemented mainly in FactSage [13] and Thermochimica [18]. Furthermore, there are several formats for thermodynamic databases including the TDB format developed by the Scientific Group Thermodata Europe (SGTE) consortium as shown for its molten salts database [19], which can be read by all software packages, and the DAT format used by FactSage [13] and Thermochimica [18]. The different database formats make it difficult to examine and update models and molten salts databases by the whole CALPHAD community.

In the present work, the MQMQA is implemented in the Python-based open-source software packages of PyCalphad [16] and ESPEI [17], making the MQMQA more accessible to the broad community. PyCalphad [16] is a Python library for designing thermodynamic models, calculating phase diagrams, and investigating phase equilibria using the CALPHAD method. ESPEI [17] is designed for evaluating model parameters with PyCalphad as the computational engine. The present implementation takes the benefits of PyCalphad and ESPEI, such as i) the prevalent scientific Python language ecosystem, ii) high throughput CALPHAD modeling based on Markov Chain Monte Carlo (MCMC) optimization, and iii) uncertainty quantification (UQ) for both models and model parameters. In addition to the capacity to read thermodynamic databases in the DAT format, the present work develops a new Extensible Markup Language (XML) format which is platform independent and programming language independent.

## 2 MQMQA and its Implementation

### 2.1 MQMQA

In the MQMQA, two sublattices are introduced to account for the first nearest neighbors (FNN) and the second nearest neighbors (SNN) interactions in a solution with quadruplets of four constituents [11,20]. In molten salt systems, the two sublattices separate the cations from the anions with two cations and two anions. Figure 1 visualizes a quadruplet (A,B)(X,Y) in a two-dimensional lattice labeled based on their chemistry [21] with three different categories, i.e., the unary quadruplets [$A_2X_2$, $B_2X_2$, $A_2Y_2$, and $B_2Y_2$], the binary quadruplets [$ABX_2$ $ABY_2$, $A_2XY$, and $B_2XY$], and the reciprocal quadruplet [ABXY]. The SNN coordination number is a model parameter related to the short-range ordering (SRO) [11]. With the SNN coordination number of A in the [ABXY] reciprocal quadruplet denoted by $Z^A_{AB:XY}$, the condition of charge neutrality for this quadruplet is maintained as follows,

$$\frac{q_A}{Z^A_{AB:XY}} + \frac{q_B}{Z^B_{AB:XY}} = \frac{q_X}{Z^X_{AB:XY}} + \frac{q_Y}{Z^Y_{AB:XY}} \qquad Eq.\ 1$$

where $q_i$ is the charges of ion $i\ (= A, B, X, or\ Y)$.

The Gibbs energy of liquid as a function of temperature ($T$) and composition (x) can be written as

$$G^{Liquid} = G^{surface} - TS^{conf} + G^{excess} \qquad Eq.\ 2$$

where $G^{surface}$ represents the reference surface energy, $S^{conf}$ the configurational entropy, and $G^{excess}$ the excess Gibbs energy of liquid. $G^{surface}$ is expressed as,

$$G^{surface} = \sum_{\substack{i=A,B}} \sum_{\substack{j=A,B \\ j \geq i}} \sum_{m=X,Y} \sum_{\substack{n=X,Y \\ n \geq m}} X_{ij:mn}\, G^o_{ij:mn} \qquad Eq.\ 3$$

where $X_{ij:mn}$ and $G^o_{ij:mn}$ are the mole fraction and the Gibbs energy per mole of the quadruplet $[ijmn]$, respectively [11]. The configurational entropy is approximately represented by,

$$-S^{conf}/R = \sum_{i=A,B} X_i \ln X_i + \sum_{m=X,Y} X_m \ln X_m + \sum_{i=A,B} \sum_{m=X,Y} X_{i:m} \ln\left(\frac{X_{i:m}}{Y_i Y_m}\right) \qquad Eq.\ 4$$

$$+ \sum_{\substack{i=A,B}} \sum_{\substack{j=A,B \\ j \geq i}} \sum_{m=X,Y} \sum_{\substack{n=X,Y \\ n \geq m}} X_{ij:mn} \ln\left(\frac{X_{ij:mn}}{\frac{f X^{e_1}_{i:m} X^{e_1}_{j:m} X^{e_1}_{i:n} X^{e_1}_{j:n}}{Y^{e_2}_i Y^{e_2}_j Y^{e_2}_m Y^{e_2}_n}}\right)$$

where $X_i$ is the mole fraction of species $i$, $X_{i:m}$ the mole fraction of the pair $i:m$, and $R$ the gas constant. The third term in the right side of Eq. 4 represents the contribution from the FNN, where $Y_i$ is the coordination equivalent site fraction and can be calculated as follows,

$$Y_i = \frac{Z_i X_i}{\sum_j Z_j X_j} \qquad \text{Eq. 5}$$

Note that the third and the fourth term of Eq. 4 are based on two different derivations from Pelton [11]. The original formalism [21], which is labeled as SUBG in the software FactSage, kept ζ (a parameter that relates how many quadruplets emanate from a specific pair) as a constant value. However, the new formalism derived by Lambotte and Chartrand [22], denominated as SUBQ in FactSage, allows ζ to be a function of composition. This change results in a modified coordination equivalent site fraction, $F_i$,

$$F_i = \sum_m X_{i:m} \qquad \text{Eq. 6}$$

and replaces the $Y_i$ term in Eq. 4. These two different formalisms affect the last term in the configurational entropy expression in Eq. 4 as well. The last term in Eq. 4 represents the contribution from the SNN, where $f$ is a factor/coefficient that equals to 1 for unary quadruplets, 2 for binary quadruplets, and 4 for reciprocal quadruplets. Note that the exponents $e_1$ and $e_2$ in the denominator of the 4th term in Eq. 4 correspond to the symbols SUBG and SUBQ used by FactSage [13], with $e_1 = 1$ and $e_2 = 1$ for the case of SUBG; and $e_1 = 0.75$ and $e_2 = 0.5$ for SUBQ. The modified expression in SUBQ gives a more accurate approximation of the quadruplet concentration when a negative value (such as -100 kJ per charge equivalent) for exchange reactions in quadruplets is applied in the model [21,22]

The excess Gibbs energy $G^{excess}$ relates to the formation Gibbs energy of the quadruplet, $\Delta g^{ex}_{quadruplet}$. For example, the following reaction forms quadruplets,

$$(A_2X_2) + (B_2X_2) = 2(ABX_2) \qquad \text{Eq. 7}$$

where $\Delta g^{ex}_{AB:X_2}$ of this equation indicates the Gibbs energy change when forming the quadruplet [ABX₂]. If $\Delta g^{ex}_{AB:X_2} = 0$, the ideal random mixing occurs. If $\Delta g^{ex}_{AB:X_2} < 0$, the reaction of Eq. 7 moves to the right, and the SRO with the [ABX₂] SNN pairs is predominant [11]. Considering multicomponent systems, Pelton [11] indicated the necessity to describe multicomponent systems with appropriate interaction parameters and taking into account what chemical groups the elements of interest would belong to. In this case the variables $\xi_{ij:X_2}$ and $\chi_{12:X_2}$ are introduced to the excess Gibbs energy expression, where the $\xi_{ij:X_2}$ is defined as follows,

$$\xi_{ij:X_2} = Y_i + \sum_k Y_k \qquad \text{Eq. 8}$$

where $k$ is the component in the $i$-$j$-$k$ system with $j$ being the asymmetrical component. The $\chi_{12:X_2}$ is defined as follows,

$$\Delta\chi_{12:X_2} = \frac{\sum_{A=1,k}\sum_{B=1,k}(X_{AB:X_2} + \sum_{Y\neq X} 0.5 X_{AB:XY})}{\sum_{A=12,k,l}\sum_{B=12,k,l}(X_{AB:X_2} + \sum_{Y\neq X} 0.5 X_{AB:XY})} \qquad \text{Eq. 9}$$

where $l$ is the component in the $i$-$j$-$l$ system with $i$ being the asymmetrical component. It should be noted that the terms regarding reciprocal quadruplets in Eq. 9 are only considered in the SUBQ category to improve interpolation contributions from the reciprocal quadruplets.

Through this approach by Pelton [11], one can study the interpolation effects based on the symmetrical and asymmetrical components in sub-ternary systems. By considering these variables, the expression of $\Delta g^{ex}_{AB:X_2}$ is obtained as follows

$$\Delta g_{AB:X_2}^{ex} = \Delta g^o + \sum_{(i+j\geq 1)} \chi_{AB:X_2}^i \chi_{BA:X_2}^j g_{AB:X_2}^{ij} \qquad Eq.\ 10$$

$$+ \sum_{\substack{(i\geq 0)\\(j\geq 0)\\(k\geq 1)}} \chi_{AB:X_2}^i \chi_{BA:X_2}^j [\sum_l \frac{g_{AB(l):X_2}^{ijk} X_{l:X}}{Y_X} (1 - \xi_{AB:X_2} - \xi_{BA:X_2})^{k-1}$$

$$+ \sum_m \frac{g_{AB(m):X_2}^{ijk} X_{m:X}}{Y_X \xi_{BA:X_2}} \left(1 - \frac{X_{B:X}}{Y_X \xi_{BA:X_2}}\right)^{k-1}$$

$$+ \sum_n \frac{g_{AB(n):X_2}^{ijk} X_{n:X}}{Y_X \xi_{AB:X_2}} \left(1 - \frac{X_{A:X}}{Y_X \xi_{AB:X_2}}\right)^{k-1} + \sum_{Y\neq X} Y_Y (1 - Y_X)^{k-1}$$

where the first two terms in the right side are binary expressions and the rest terms describe ternary interactions with symmetrical and asymmetrical considerations for the components being mixed. The first of the three ternary interactions represents the scenario when either component $k$ is the asymmetrical component or when all components are symmetrical. The second term is when the species B is the asymmetrical component in the system. The third term is when the species A is the asymmetrical component in the system. The final term represents a reciprocal ternary interaction parameter to describe the interaction between [AB:X2] quadruplet when component Y is present.

## 2.2  Implementation of MQMQA into PyCalphad and ESPEI

Figure 2 shows the flowchart regarding the implementation of MQMQA into PyCalphad, where the new features of the present efforts are highlighted by green and the pre-existing features are highlighted by blue. The first step of implementation is reading in the database in different formats, i.e., the DAT file in the FactSage/ChemSage format [13,18], or the TDB format in Thermo-Calc [12], OpenCalphad [15], Pandat [14], and PyCalphad/ESPEI [16,17], or the XML format proposed in the present work.

To support a unified, on-disk representation of CALPHAD-type databases that avoids implementing software-specific extensions to the *de facto* standard TDB and DAT formats, we propose a new database format using XML. Unlike the existing domain specific languages used in CALPHAD databases that require each software tool to develop bespoke implementations of non-standard, *ad hoc* syntaxes, XML is a well-established, standardized markup language that is human- and machine-readable. Virtually all programming languages have XML parsers implemented in their standard libraries or in common third-party libraries. To describe the domain-specific aspects, we use a declarative Relax NG-based schema [23] that uses XML syntax. The XML parser for PyCalphad is implemented as a separate plugin package called "PyCalphad-xml" [24], that provides function handles that are used by PyCalphad database objects to read, write, and validate XML databases using the Relax NG schema provided by the PyCalphad-xml package.

The second step of implementation is to perform equilibrium calculations (*cf.*, Figure 2) using the MQMQA and any other models for specific phases. Since ESPEI uses PyCalphad as its computational engine, the implementation of MQMQA into PyCalphad makes it possible for ESPEI to perform CALPHAD modeling using MQMQA. UQ of model parameters and MCMC for database development are then straightforward; see brief discussion in Sec. 1 as well as the discussion in the literature [16,17].

## 3 Demonstration of Equilibrium Calculations using PyCalphad

Using the NiF$_2$-KF database modeled by Ocadiz-Flores et al. [25] in terms of MQMQA and the FactSage software, we validate the implementation of MQMQA in PyCalphad [16,26] by performing equilibrium calculations. For the convenience of users, there is also a Jupyter Notebook as Supplemental Material to demonstrate the details of calculations.

Figure 3 shows the calculated Gibbs energy, entropy, and enthalpy of NiF$_2$-KF at 1600 K, which is in the liquid phase region using MQMQA in terms of PyCalphad. For the purpose of comparison, the Gibbs energies are calculated using PyCalphad [16], Thermochimica [18], and FactSage [13]. Figure 3 shows that the calculated results of Gibbs energy, entropy, and enthalpy from PyCalphad agree well with those from Thermochimica and FactSage, indicating the successful implementation of MQMQA into PyCalphad. Figure 4 shows the Gibbs energy surface of NiF$_2$-KF liquid at 1600 K by sampling of the internal degrees of freedom (quadruplet fractions in the present case) [27]. It presents the overall surface at each composition consists of different configurations of quadruplet fractions and the respective energies that belong to them.

Figure 5 is the calculated pseudobinary phase diagram of KF-NiF$_2$ in comparison with experimental data used to model this system by Ocadiz-Flores et al. [25]. In addition, the calculated KF-NiF$_2$ phase diagram by both PyCalphad and Thermochimica is shown in the Supplemental Figure S1. It is seen that the phase diagram is correctly reproduced by PyCalphad and is in a good agreement with experimental data and the predictions by Thermochimica (*c.f.,* Figure 5 and the Supplemental Figure S1). Figure 6 shows the calculated quadruplet fractions as a function of NiF$_2$ mole fraction, x(NiF$_2$), at 1600 K for the three quadruplets of [K$_2$F$_2$], [Ni$_2$F$_2$], and [KNiF$_2$] in the NiF$_2$-KF system. The overall distribution of the [KNiF$_2$] quadruplet skews toward x(NiF$_2$) = 0.38, which relates to the composition with the maximum SRO (x(NiF2) = 0.333) based on the coordination numbers chosen in MQMQA.

## 4  Demonstration of thermodynamic modeling with UQ using ESPEI

Once the user provides (i) a thermodynamic database with estimated values of the adjustable parameters and (ii) input datasets describing thermochemical and phase boundary data, model parameters in the database can be simultaneously optimized using the MCMC method as implemented in ESPEI [17]. The statistical distributions of the model parameters are evaluated from the samples during MCMC optimization based on the Metropolis criteria [17]. In the present work, the excess Gibbs energy of liquid (i.e., $\Delta g^{ex}_{AB:X_2}$ in Eq. 10) were chosen as model parameters and remodeled to quantify uncertainty using a 95% uncertainty interval (or Bayesian credible intervals containing 95% of the invariant samples). During the CALPHAD modeling process by ESPEI/PyCalphad, the Markov chain values can be tracked and the MCMC steps were performed until all model parameters converged. Note that 8 chains were selected for each model parameter with a total of 800 MCMC iterations performed. The Supplemental Material includes a Jupyter Notebook for the details of CALPHAD modeling with UQ.

Similarly to those in Sec. 3, the KF-NiF$_2$ system modelled by Ocadiz-Flores et al. [25] were selected for demonstration; where the input datasets in the JavaScript Object Notation (JSON) files with phase boundary data were used as constraints during the modeling process, see the experimental data in Figure 5 represented by symbols or the JSON files provided in the Supplementary Material. Figure 7 shows the evolution of Markov chains during MCMC iterations indicated by the log-probability changes for all the chains as a function of iteration. After around 20 MCMC iterations, the chains are roughly converged to the similar log-probabilities around -$1.48 \times 10^3$, indicating the convergence of MCMC optimization of model parameters. After a total of 800 MCMC steps, the modelled parameters values as well as the initial parameters by Ocadiz-Flores et al. [25] are listed in Table 1; showing no significant change after the present MCMC interactions since the present work is only for the purposes of demonstration and the initial parameters were already modelled by Ocadiz-Flores et al. [25].

The uncertainty of thermodynamic properties can be quantified using the UQ method after CALPHAD modeling by MCMC [28]. UQ relies on PyCalphad for predicting thermodynamic properties of interest and ESPEI for Bayesian samples to leverage the distribution of model parameters and estimate uncertainties based on the estimated Gaussian distribution of input data [28]. For the purpose of simplicity, only the results in last MCMC step were used for the present UQ study. Note that more MCMC steps cannot improve the UQ results after the convergence of MCMC. As an example, Figure 8 shows the uncertainty of phase fractions as a function of temperature with x($NiF_2$) = 0.32 for the KF-$NiF_2$ system. The plot presents two different invariant reactions (KF+$NiK_2F_4$ → liquid and $NiK_2F_4$ → $NiKF_3$ + liquid) and two two-phase regions (KF + $NiK_2F_4$ and $NiKF_3$ + liquid) before complete melting of the system. It shows that the eutectic reaction of 1075 K has the lowest uncertainty of the invariant reactions, and the liquid + $NiKF_3$ region has the highest uncertainty based on the shaded region plotted. These results are expected since the experimental data for liquidus are more scattered than the measured invariant reactions at 1075 K (see experimental data points shown in Figure 5). Figure 8 also shows that the uncertainties of phase fraction increase from 1200 K to 1400 K, where the peritectic reaction

"NiK$_2$F$_4$→ NiKF$_3$ + liquid" occurs (see also the phase diagram in Figure 5), and then, the NiKF$_3$ + liquid two-phase region appears. The shadow regions of NiKF$_3$ and liquid phase indicate that the liquidus temperature is with uncertainties from 1300 K to 1400 K at x(NiF$_2$) = 0.32. These uncertainty results are expected since data scattering occurs in different experiments [25,29,30] when measuring these invariant reaction and liquidus.

As another example, Figure 9 displays the uncertainty of the Ni activity as a function of composition x(Ni) at 1800 K. The uncertainties of the activity with x(Ni) = 0.15 - 0.25 are large. There is no experimental data of activity of Ni at 1800 K. However, the model parameters were optimized based on phase boundary data shown in Figure 5, where more scattered experimental data points are found in the range of x(Ni) = 0.16 - 0.22 (i.e., x(NiF2) around 0.38 to 0.6). It results in larger uncertainty of model parameters in this composition range, leading to larger uncertainty of activity calculated from the MCMC process. These results (Figure 8 and Figure 9) demonstrate application of UQ to study phase equilibria and thermodynamic properties of solutions modeled with MQMQA for the first time.

## 5   Summary

In the present work the modified quasichemical model in the quadruplet approximation (MQMQA) is implemented into the Python-based open-source software PyCalphad. It is now possible to use the open-source software ESPEI for database development using MQMQA with uncertainty quantification (UQ). A new database format based on Extensible Markup Language (XML) is proposed to efficiently represent MQMQA models. Using the KF-NiF$_2$ system as an example, the capacities of the implemented MQMQA to perform equilibrium thermodynamic calculations and develop new UQ-enabled databases are demonstrated. The present predictions agree well with experiments and calculated results available in the literature, indicating the successful implementation of MQMQA into PyCalphad/ESPEI. The present implementation makes the MQMQA more accessible to the broad community.

# 6 Acknowledgements

The authors at Penn State acknowledge the financial support by the U.S. Department of Energy (DOE) via Award No. DE-NE0008945. Part of this work was performed under the auspices of the U.S. DOE at Lawrence Livermore National Laboratory under Contract No. DE-AC52–07NA27344. Part of this work was carried out at the Jet Propulsion Laboratory (JPL), California Institute of Technology, under a contract with the National Aeronautics and Space Administration (80NM0018D0004). Partial support from the Caltech/JPL Innovative Software Initiative program is also acknowledged. B.B. acknowledges support from a NASA Space Technology Research Fellowship (NSTRF 80NSSC18K1168).

**Figures and Figure Captions:**

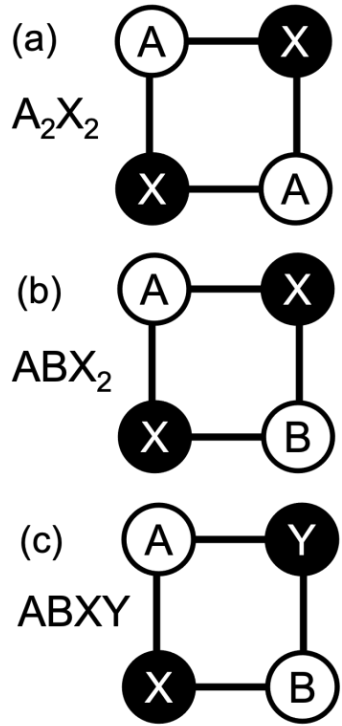

Figure 1. Examples of three categories of quadruplets: (a) $A_2X_2$ unary quadruplet, (b) $ABX_2$ binary quadruplet, and (c) ABXY reciprocal quadruplet. This figure is reproduced based on Pelton et al.'s work [21].

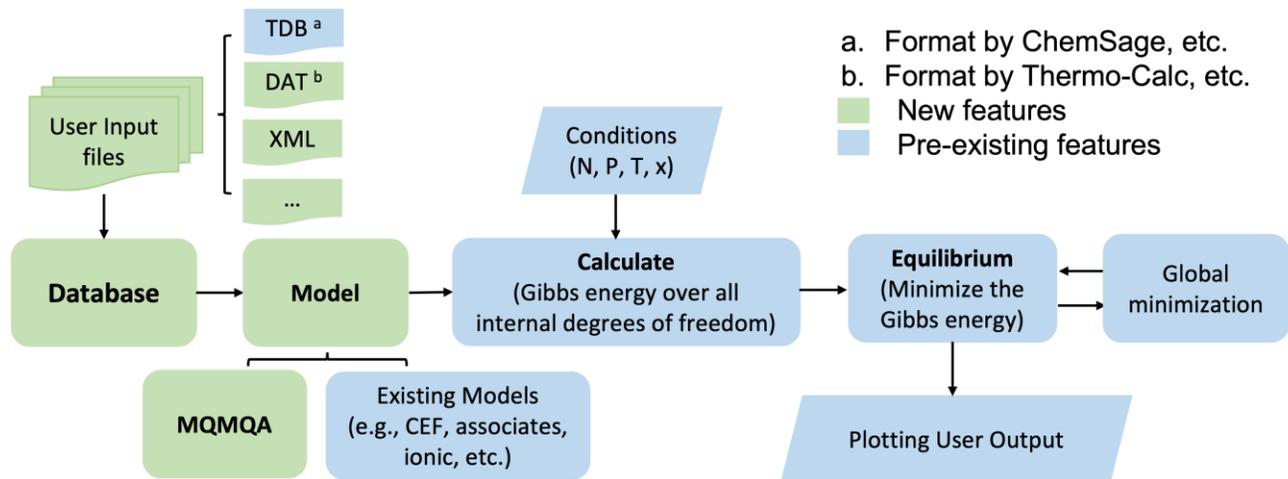

*Figure 2.* Flowchart that describes the implementation of MQMQA into the PyCalphad software [16].

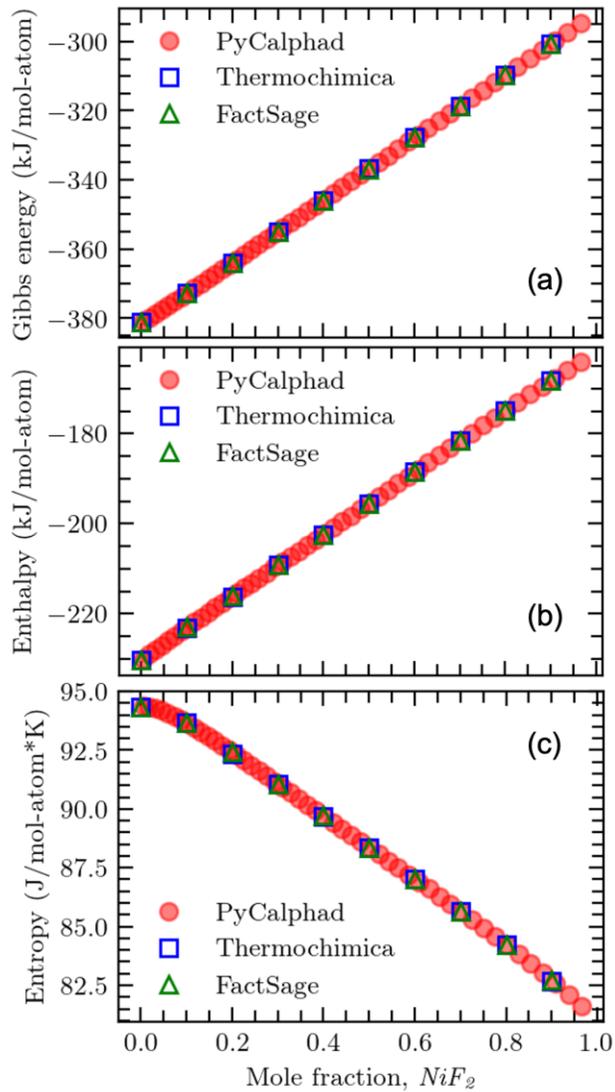

*Figure 3.* Calculated results of (a) Gibbs energy, (b) enthalpy, and (c) entropy of the liquid phase in the KF-NiF$_2$ system as a function of x(NiF$_2$) at 1600 K. The solid red circles are the results calculated by PyCalphad, the blue open squares by Thermochimica, and the green open triangles by FactSage. These calculations are all based on Ocadiz-Flores' database [25].

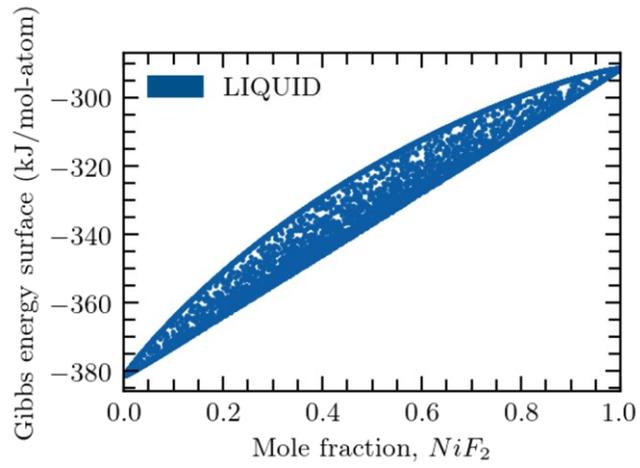

*Figure 4.* Gibbs energy surface of the liquid phase in the KF-NiF$_2$ system at 1600 K calculated using PyCalphad from Ocadiz-Flores' database [25].

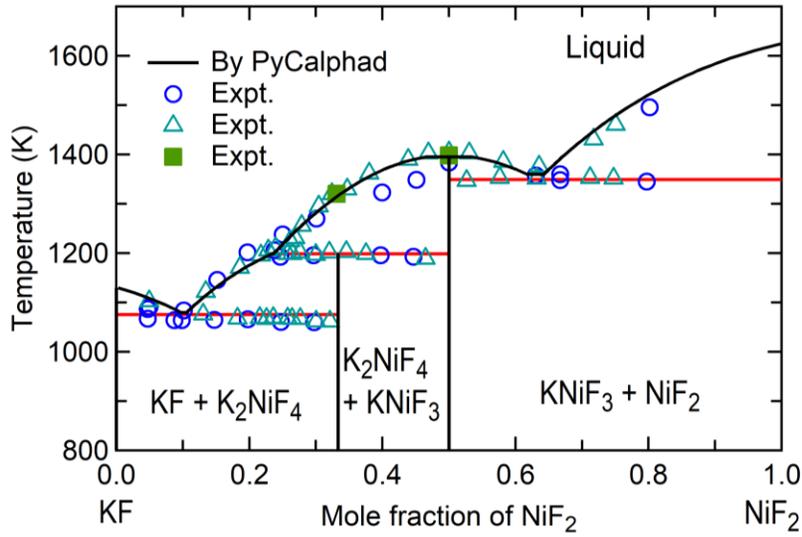

*Figure 5.* Calculated KF-NiF$_2$ phase diagram by PyCalphad using Ocadiz-Flores' database [25] in comparison with experimental data points (the symbols) summarized by Ocadiz-Flores et al. [25].

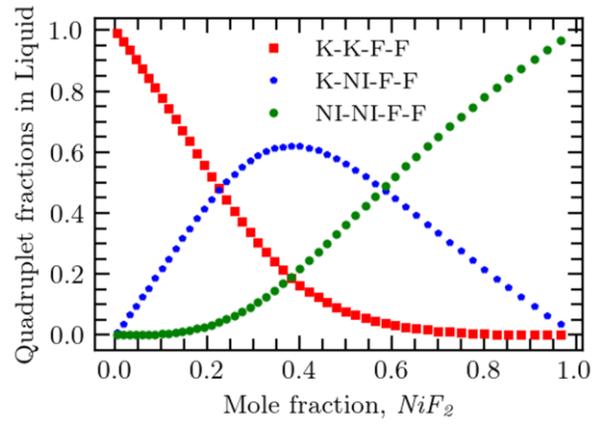

*Figure 6.* Quadruplet fractions as a function of x(NiF$_2$) at 1600 K in the liquid phase using Ocadiz-Flores's database [25] and calculated by PyCalphad.

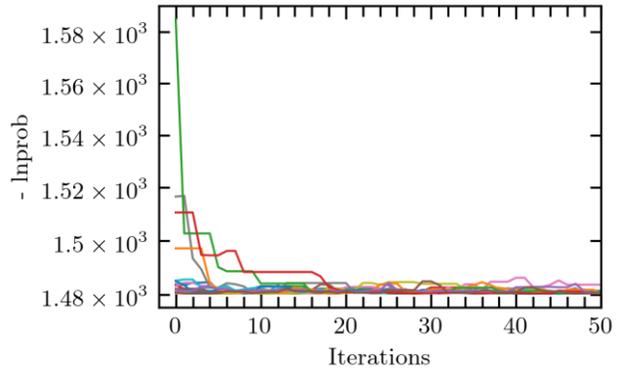

*Figure 7.* Convergence of chains in first 50 MCMC iterations and 800 iterations were used for MCMC. The y-axis label of '- lnprop' represents the logarithm of the probability with a negative sign to make the value positive.

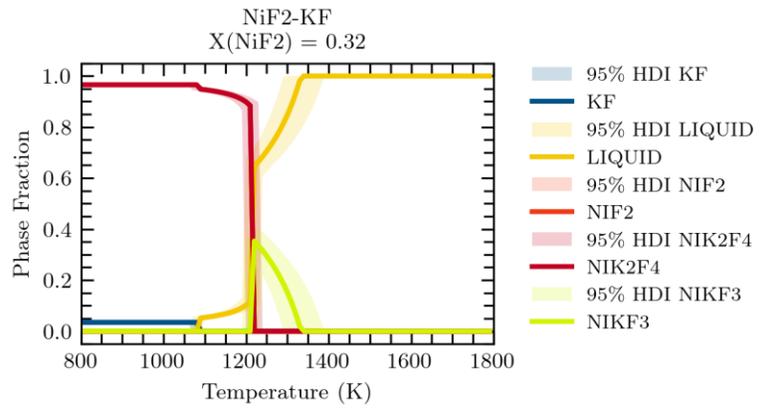

*Figure 8.* Uncertainties of phase fractions as a function of temperature in the KF-NiF$_2$ system with x(NiF$_2$) = 0.32 predicted after the MCMC. The shaded area represents the 95% high density interval (HDI).

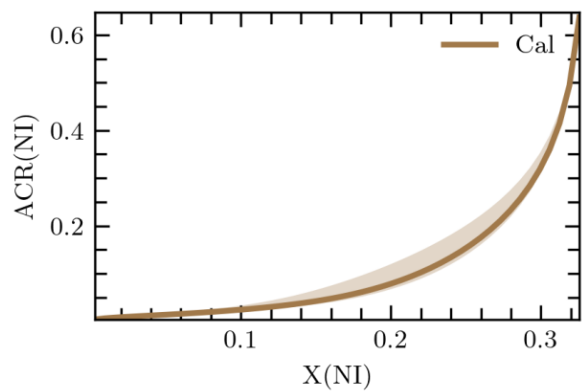

*Figure 9.* Uncertainties of activity of Ni, ACR(NI), as a function of composition x(Ni) in the KF-NiF$_2$ system at 1800 K predicted after the MCMC. The shaded area represents the 95% high density interval (HDI).

**Tables**

*Table 1.* Gibbs energy parameters of liquid in the KF-NiF$_2$ system after 800 MCMC iterations, compared with the initial values from Ocadiz-Flores et al. [25].

| Parameter values | Reference |
|---|---|
| $\Delta g^{ex}_{KNi:F_2} = -17062.4 - 13654.3\chi_{NiK/F_2}$ | This work after 800 MCMC iterations |
| $\Delta g^{ex}_{KNi:F_2} = -17573 - 15899\chi_{NiK/F_2}$ | Ocadiz-Flores et al. [25] |

**Supplemental Material**

**Note that the files in the following Supplemental Tables can be found also in:**
https://doi.org/10.5281/zenodo.6471272

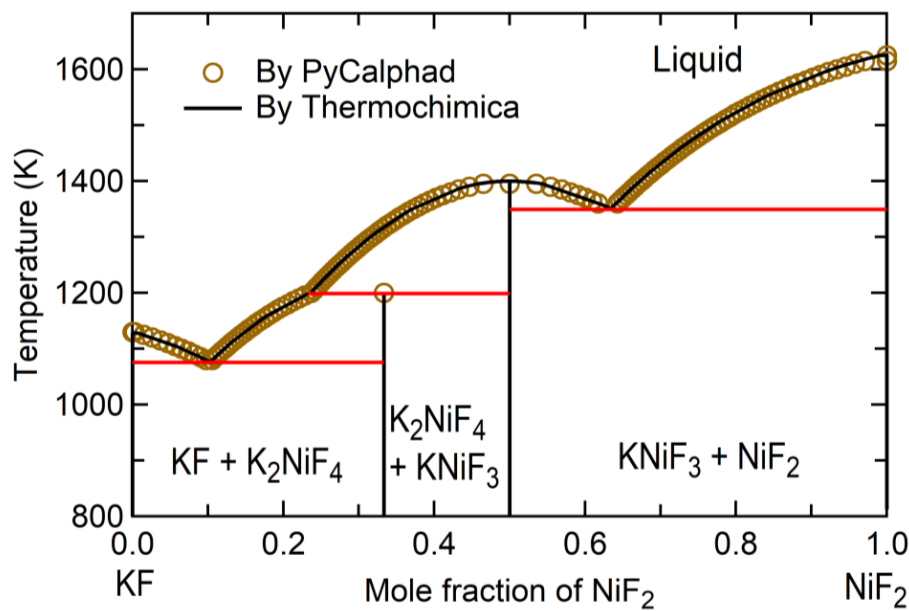

*Figure S1*. Predicted phase diagrams of the KF-NiF$_2$ system by PyCalphad (symbols) and Thermochimica (lines) in terms of the MQMQA and the database developed by Ocadiz-Flores, et al. [25].

**Table S1**. List of folder and file names.

| Folder or file name | Notes |
|---|---|
| pycalphad | A folder with Jupyter notebook example and files for using PyCalphad for equilibrium calculations. |
| ESPEI | A folder with Jupyter notebook example and files for using ESPEI for modeling optimization. |
| Installation.md | File with instruction for installation of PyCalphad/ESPEI |

**Table S2**. Explanation of the Jupyter notebook "MQMQA-DEMO.ipynb" in folder "pycalphad".

| Item | Comments and notes | |
|---|---|---|
| Purpose | To perform equilibrium calculations for system modelled with MQMQA using PyCalphad, including phase diagrams, Gibbs energy, etc. | |
| Input files | Ocadiz_combined.dat | KF-NiF$_2$ database from Ocadiz-Flores et al., DOI: 10.1016/j.jct.2018.01.023 |
| | MSTDB_test.dat | MSTDB from T. Besmann et al., DOI: 10.1016/j.jnucmat.2022.153631 |
| Output files | MSTDB.xml | MSTDB in XML format. |

**Table S3.** Explanation of the Jupyter notebook "MCMC-DEMO.ipynb" in folder "ESPEI".

| Item | Comments and notes | |
|---|---|---|
| Purpose | To perform optimization for systems with MQMQA using ESPEI, where the databases are in XML format. | |
| Input files | input_data | Subfolder containing experimental data of KF-NiF$_2$ summarized by Ocadiz-Flores et al., DOI: 10.1016/j.jct.2018.01.023. |
| | ESPEI_demo.yaml | Setting file for EPSEI running in yaml format. |
| | Ocadiz_combined.dat | KF-NiF$_2$ database from Ocadiz-Flores et al., DOI: 10.1016/j.jct.2018.01.023. |
| | Ocadiz_combined.xml | Ocadiz_combined.dat converted into XML format. Note that the adjustable parameters (starting from VV) should be set before running MCMC; see details in https://espei.org. |
| | Phases.json | JSON file to specify models of phases in the database. |
| Output files | MCMC_demo.xml | Output database after MCMC optimization. |
| | lnprob_demo.npy | File recording probability of chains during MCMC. |
| | trace_demo.npy | Trace file for MCMC. |